\documentclass{ws-procs9x6}
\def\etal {{\it et al.}}
\begin{document}
\title{Constraints on Possible Monopole-Dipole Interactions of WISPs from the Transverse Relaxation Time of Polarized $^3$He Gas}
\author{Changbo Fu$^{a,b}$, Thomas R. Gentile$^{a}$, and William M. Snow$^{b}$}

\address{$^{a}$National Institute of Standards and Technology, Gaithersburg, MD 20899\\
$^{b}$Center for Exploration of Energy and Matter, Indiana Univ., Bloomington, IN 47408}

%\author{Thomas R. Gentile}

%\address{National Institute of Standards and Technology, Gaithersburg, MD 20899\\E-mail:thomas.gentile@nist.gov}

%\author{William M. Snow}
%\address{Center for Exploration of Energy and Matter, Indiana Univ., Bloomington, IN 47408\\ E-mail:wsnow@indiana.edu}

\begin{abstract}

Various theories beyond the Standard Model predict new particles with masses in the sub-eV range with very weak couplings to ordinary matter. 
%\textit{ A $P$-odd and $T$-odd interaction between polarized and unpolarized matter proportional to ${\vec{s}} \cdot {\vec{r}}$ is one such possibility. Such an interaction involving a scalar coupling $g_{s}$ at one vertex and a pseudoscalar coupling $g_{p}$ at the other vertex (sometimes referred to as a monopole-dipole interaction in the literature) can be induced by the exchange of spin-$0$ particles between two vertexes.}
A $P$-odd, $T$-odd, spin-dependent interaction between polarized and unpolarized matter is one such possibility. Such a monopole-dipole interaction can be induced by the exchange of spin-$0$ particles.
The presence of a possible monopole-dipole interaction between fermion spins and unpolarized matter would cause an decreased transverse spin relaxation time $T_{2}$ for a confined gas of polarized nuclei.
By reanalyzing previously existing data on the spin relaxation times of polarized $^3$He in gas cells with pressure in the millibar range and applying the well-established theory of spin relaxation for magnetic field gradients to gradients in a possible monopole-dipole field, we present new laboratory constraints on the strength and range of such an interaction. These constraints represent to our knowledge the best limits on such interactions for the neutron with ranges between $0.01$ cm and 1 cm.

\end{abstract}

\bodymatter

\section{Introduction}

The possible existence  of new particles beyond the Standard Model  with masses in the sub-eV range and very weak couplings to ordinary matter is starting to attract increased attention\cite{WISP-NPB}. These possible particles are now starting to be  referred to in the literature as WISPs (Weakly-Interacting sub-eV Particles)\cite{WISP-NPB, Jaeckel10}.  A $P$-odd and $T$-odd interaction between
polarized and unpolarized matter proportional to ${\vec{s}} \cdot {\vec{r}}$ (spin and distance) is one such possibility. An interaction involving a scalar coupling $g_{s}$ at one vertex and a pseudoscalar coupling $g_{p}$ at the other vertex (sometimes referred to as a monopole-dipole interaction in the literature) can be induced by the exchange of spin-$0$ particles between two vertexes. The axion is one possible example of such a particle, and many laboratory experiments and astrophysical observations have searched for it\cite{Review-Astro-Axion-19901}.

The monopole-dipole interaction potential between nucleon spins and unpolarized matter is given by\cite{Axion.New-Force.Moody.PRD}
\begin{equation}\label{eq.potential.org}
V=\hbar^{2}g_{s}g_{p}\frac{\hat{\mathbf{\sigma}}\cdot\mathbf{\hat{r}}}{8\pi m_n}
\left(
\frac{1}{r\lambda}+\frac{1}{r^{2}}
\right)
e^{-r/\lambda},
\end{equation}
where $m_n$ is the mass of the nucleon at the polarized vertex, $\hbar$ is the reduced Planck constant, $\hbar\hat{\sigma}/2$ is the fermion spin, $\lambda=\hbar/m_ac$ is the range of the interaction, $m_a$ is the mass of axion, and $\hat{\mathbf{r}}={\mathbf{r}}/r$ is the unit vector between the dipole and the mass.

Measurements of the longitudinal relaxation time $T_{1}$ of polarized $^{3}$He were recently used to constrain the interaction strength and range of such a possible monopole-dipole interaction coupling to neutrons\cite{Serebrov2009423,Yuri-T1-3He}. The method takes the advantage of the fact that the motion of a polarized species through a possible monopole-dipole field gradient will cause the polarization to decay. In Ref. [\refcite{Serebrov2009423}] the author constrained the monopole-dipole interaction  by using the  depolarization rate of ultracold neutrons (UCN) in a material trap. In Ref. [\refcite{Yuri-T1-3He}] the author employed the longitudinal relaxation time $T_{1}$ of polarized  $^3$He gas at pressures of a few bar to set a constraint on the monopole-dipole interaction strength.

In this work we constrain the strength of a possible monopole-dipole interaction of neutrons by using previous measurements of the transverse spin relaxation time $T_{2}$ of  polarized $^3$He gas at pressures of a few mbar. We find that this is the best constraint to our knowledge in the distance range from $0.01$ cm to $1$ cm for a monopole-dipole interaction involving neutrons. This constraint could be greatly improved in dedicated experiments.

\section{Spin Relaxation in a Low Pressure Cell}

Consider a dipole in a uniform magnetic field $\textbf{B}_0$ along the z axis and a very large mass block of thickness $d$ at position $z$. If a monopole-dipole interaction exists,  the potential has the form\cite{Serebrov2009423}
\begin{equation} \label{eq.potential.plane}
V(\mathbf{r})=\frac{g_{s}g_{p}N\hbar^{2}\lambda}{4 m_n}e^{-z/\lambda}(1-e^{-d/\lambda}),
\end{equation}
where $N$ is the nucleon density of the matter. Consider a spherical cell with radius $R$ containing low density polarized gas. The polarized gas in the cell sees both the monopole-dipole field $V$ and the normal magnetic field $\textbf{B}_0$. If $R\gg \lambda$, the average variation of the monopole-dipole interaction potential over the whole cell can be written as
\begin{equation}
\langle\Delta V\rangle\approx \frac{4\pi R^2\lambda}{\frac{4}{3}\pi R^3}\frac{g_{s}g_{p}N\hbar^{2}\lambda}{4 m_n}(1-e^{-d/\lambda}).
\end{equation}

Because both the monopole-dipole interaction and the magnetic interaction $\vec{\mu} \cdot \vec{B}$ are of the form $\vec{s} \cdot \vec{\hat{r}} f(r)$,  one would expect an extra contribution to the spin relaxation due to the monopole-dipole interaction.\cite{McGregor-T2}
In analogy with a magnetic field gradient, the transverse relaxation rate induced by the monopole-dipole field gradient can be written as\cite{Pines-1955}
\begin{equation}
\frac{1}{T_2^{'}}=(\delta\omega)^2\,\tau_c ,
\end{equation}
where $\tau_c$ is the correlation time for the relaxation mechanism under consideration and $\hbar\,\delta \omega$ is the interaction energy in the  monopole-dipole interaction field. The decrease in the transverse relaxation rate $\delta\omega$ by the factor $\delta\omega\,\tau_c$ due to the motion of the dipole is known as motional narrowing. 
Using the correlation time for the random motion of polarized atoms in an ideal gas
$\tau_c\approx\frac{R^2}{2D}$,
where $D$ is diffusion constant, the transverse relaxation rate from the monopole-dipole interaction becomes
\begin{equation}
\frac{1}{T_2^{'}}\approx \left[
\frac{3\lambda^2\, g_sg_pN \hbar^2}{4R\,m_n}
(1-e^{-d/\lambda})
\right]^2
\frac{R^2}{2D}.
\end{equation}

The relaxation rates from independent processes add, yielding a total rate $1/T_2$ giving by 
\begin{equation}
\frac{1}{T_2}=\frac{1}{T_2^{(w)}}
+\frac{1}{T_2^{(\partial B)}}
+\frac{1}{T_2^{(dd)}}
+\frac{1}{T_2^{'}},
\end{equation}
where we list the dominant sources: wall collision ($1/T_2^{(w)}$), inhomogeneity of the external magnetic field ($1/T_2^{(\partial B)}$), relaxation related to the dipole-dipole interaction ($1/T_2^{(dd)}$), and inhomogeneity of the monopole-dipole field ($1/T_2^{'}$). By assuming that the difference between the measured $T_{2}$ and that calculated from theory is due to the monopole-dipole interaction, one can set a limit on $g_sg_p$ using
\begin{equation}
g_sg_p\le \frac{4\,m_n}{3N\hbar\lambda^2} 
\sqrt{\frac{2D}{T_2}} \frac{1}{(1-e^{-d/\lambda})},\ \ (\lambda\ll R).
\end{equation}

\section{Monopole-dipole interaction coupling constant constraints on the neutron from previously existing measurements of $T_{2}$ spin relaxation measurements of $^{3}$He}

We now use these relations to set limits on the monopole-dipole interaction on the neutron spins in polarized $^{3}$He. In Ref. [\refcite{long-T2-Gemmel-low-P}], $T_2$ measurements are reported for a spherical aluminosilicate glass cell of radius $R=3$ cm filled with $^3$He to a pressure of $4.5$ mbar. The transverse relaxation time was measured to be $T_2=(60.2\pm0.1)$ h. The diffusion constant is $D= 0.04$ m$^2$/s. The nucleon density of the aluminosilicate glass is $N= 1.31\times 10^{30}$ m$^{-3}$. Assuming a cell wall thickness  of $1.5$ mm, typical for glass cells used in magnetometry work of this type, from the relations above the limit on a monopole-dipole interaction involving the neutron is\begin{eqnarray}
g_s^{(n)}g_p^{(n)}\leq \frac{9.8\times 10^{-23}}{\lambda^{2}(1-e^{-0.15/\lambda})},\,\ \lambda[{\rm cm}]\ll3 \ {\rm cm}.
\end{eqnarray}

In Ref. [\refcite{McGregor-T2}], a 13.1 mbar, 4.96-cm-diameter spherical cell was used to study transverse spin relaxation times for polarized $^{3}$He  in measured magnetic field gradients. The diffusion constant of $^3$He in the cell is $D=0.0137$ m$^2$/s. They measured $T_2$ in a nonmagnetic building with low magnetic field gradients to be 14.86 h. Assuming the cell wall nucleon density was $1.3\times10^{30}$ m$^{-3}$ and the wall thickness is 2 mm, the constraint on the monopole-dipole interaction strength is 
\begin{eqnarray}
g_s^{(n)}g_p^{(n)}\leq \frac{1.2\times 10^{-22}}{\lambda^{2}(1-e^{-0.2/\lambda})},\,\ \lambda[{\rm cm}]\ll 5\ {\rm cm},
\end{eqnarray}
which is very close to the constraint above.

\begin{figure}
\psfig{file=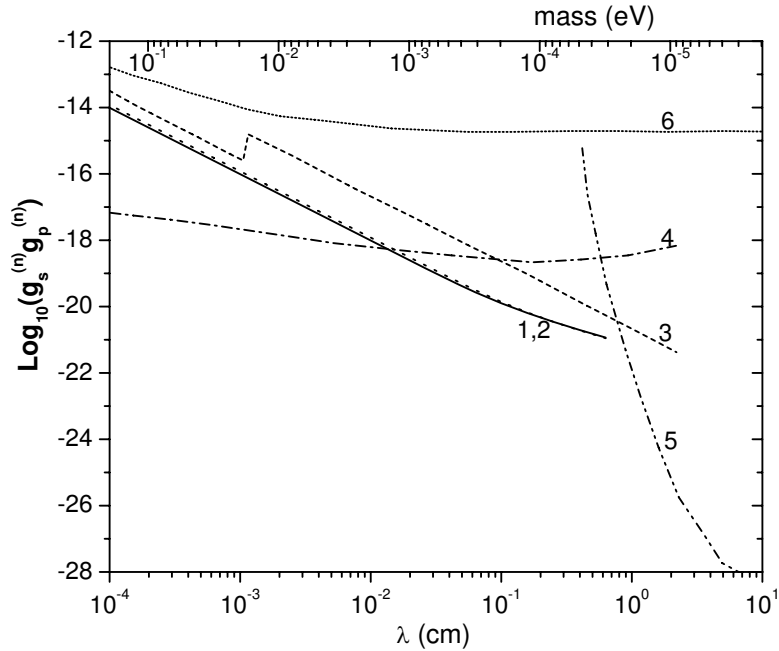,width=4.1in}
\caption{Constraints on the monopole-dipole  coupling strength $g_s^{(n)}g_p^{(n)}$: 1. Solid line, this work, by reanalyzing the data in Ref. [\refcite{long-T2-Gemmel-low-P}]; 2. Dot line, this work, by reanalyzing the data in Ref. [\refcite{McGregor-T2}]; 3. Short dash line, from Ref. [\refcite{Serebrov2009423}]; 4. Dash-dot line, from Ref. [\refcite{Yuri-T1-3He}]; 5. Dash-dot-dot line, from Ref.  [\refcite{Youdin-HgCs}]; 
6. Short-dot line, from Ref. [\refcite{Bab-UCN}].
%7. Short-dash-dot line, this work, see the text for details.
 }
\label{fig.result}
\end{figure}

Finally, we note that there are unpublished reports of even longer $T_{2}$ (140 h) for polarized $^{3}$He gas in glass cells\cite{TI-T2-result}. 
Unfortunately the pressure and the wall material information are not shown in the reports, hence we do not use these results to set a limit.
%With reasonable assumptions of $R=5$ cm, $N=1.3\times10^{30}$ m$^{-3}$, wall %thickness 2 mm, and $P=15$ mbar, a constraint can be obtained Fig. %\ref{fig.result}(7).

From measurements and theoretical calculations\cite{Friar90}, it is known that the polarization of the $^{3}$He nucleus is dominated by the neutron polarization, with only a small contribution from orbital motion and other effects. Therefore these limits are to a good approximation interpretable simply as constraints on a neutron monopole-dipole interaction. With the inclusion of constraints from $T_{2}$ measurements in other nuclei, such as $^{129}$Xe, it would be possible to place separate constraints on the monopole-dipole interactions of neutrons and protons. This work is in progress.

\section{Summary}

In this work we present constraints on the strength and range of a possible monopole-dipole interaction involving the neutron.
These new laboratory limits, set by reanalysis of previous measurements of the transverse spin relaxation time $T_{2}$ in polarized $^{3}$He cells, are the best in existence for the neutron to our knowledge in the range $0.01$ cm to 1 cm. Constraints on the monopole-dipole interaction using this method can be significantly improved in dedicated experiments.

This work was supported in part by the National Science Foundation under award PHY-0116146.

%\bibliography{../5forceCitations}

%\bibliographystyle{apsrev}

\end{document}